\documentclass{elsart}
	
	\newif\ifpdf
	\ifx\pdfoutput\undefined
	\pdffalse    
	\else
	\pdfoutput=1 
	\pdftrue
	\fi
	
	\ifpdf
	\usepackage[pdftex]{graphicx}
	\else
	\usepackage{graphicx}
	\fi
	\usepackage{amsfonts}

\begin{document}
	\ifpdf
	\DeclareGraphicsExtensions{.pdf}
\else
	\DeclareGraphicsExtensions{.eps}
\fi

\begin{frontmatter}

	\title{Nambu--Goto Strings from $SU(\, N\,)$ Born-Infeld model}

\author[Trieste]{Stefano Ansoldi}
\thanks{e-mail address: ansoldi@trieste.infn.it},
\author[Atlanta]{Carlos Castro}
\thanks{e-mail address: castro@ctsps.cau.edu},
\author[bgu]{E. I. Guendelman}
\thanks{e-mail address:guendel@bgumail.bgu.ac.il },
\author[Trieste]{Euro Spallucci}
\thanks{e-mail address: spallucci@trieste.infn.it}

\address[Trieste]{Dipartimento di Fisica Teorica,
				  Universit\`a di Trieste
		          and INFN, Sezione di Trieste}
\address[Atlanta]{Center for Theoretical Studies of Physical Systems,
				  Clark Atlanta University, Atlanta, GA.30314 }
	\address[bgu]{Physics Department, Ben Gurion University, Beer Sheva,
	Israel}

	\begin{abstract}
	The spectrum of quenched Yang-Mills theory in the large--$N$ limit 
	displays strings and higher dimensional extended objects. The
	 effective dynamics of  string-like excitations is encoded into
	 area preserving Schild action. In this letter, we bridge the gap
	 between $SU(\,N\,)$ gauge models and fully reparametrization invariant
	 Nambu--Goto string models by introducing an extra matrix degree of 
	 freedom in the Yang-Mills action.
	 In the large--$N$ limit this matrix variable becomes the world-sheet
	 auxiliary field  allowing a smooth transition between the Schild 
	 and Nambu--Goto strings. The new improved matrix model we propose here
	 can be extended to $p$-branes provided we enlarge the dimensionality
	 of the target spacetime.
	 \end{abstract}
	\end{frontmatter}
	
	In the \textit{large--$N$} limit $SU(N)$ Yang-Mills gauge theories
	display string-like excitations \cite{uno}. The effective dynamics of 
	these one-dimensional objects is described by a Schild action which is
	invariant under area preserving reparametrizations only. This
	result allowed to establish a relationship between $SU(\, \infty\,)$
	and symplectic transformations: $\sigma^m\to \sigma^{\prime\,m}=
	\sigma^{\prime\, m}(\sigma)$,
	$\vert\, {\partial\sigma^\prime \over \partial\sigma}\,\vert=1$, but 
	not between $SU(\, \infty\,)$ and the group of general reparametrization
	encoded into the Nambu--Goto action.\\
	In this letter we are going to show how to recover the
	reparametrization invariant Nambu--Goto action form the large--$N$ limit
	of a $SU(N)$ gauge invariant action.\\
	As a starting point, let us consider
	an action smoothly interpolating between the (~area preserving~)
	Schild action and the fully (~reparametrization invariant~) Nambu--Goto
	action can be written by introducing an auxiliary world--sheet field
	$\Phi(\, \sigma\,)$ \cite{oda}, :
	 
	 \begin{equation}
	I\left[\, \Phi\ , X\,\right]\equiv \frac{\mu_0}{2}\, \int_\Sigma d^2\sigma\,
	\left[\, {\mathrm{det}(\gamma_{mn})\over 
	\Phi(\,\sigma\,)} + \Phi(\, \sigma\,)\,\right]
	\label{fix}
	\end{equation}
	
	where, $\gamma_{mn}\equiv \eta_{\mu\nu}\, \partial_m\, X^\mu\,
	\partial_n\, X^\nu$ is the \textit{induced metric} on the string
	Euclidean world-sheet $x^\mu= X^\mu(\,\sigma\,)$,
	$\mathrm{sign}(\gamma_{mn})=(+\ ,+)$; finally,
	$\mu_0\equiv 1/2\pi\alpha^\prime$ is the string tension. We assigned 
	the following dimension (~in natural units~) to the various quantities 
	in (\ref{fix}):
	 \begin{equation}
	\left[\, \sigma^m\,\right]= \mathrm{length} \ ,\qquad \left[\,
	X^\mu\,\right]=\mathrm{length}\ ,\qquad \left[\, \Phi\,\right]=1
	\end{equation}

	For later convenience, we recall the relation between
	$\mathrm{det}(\gamma_{mn})$ and the world manifold Poisson Bracket:
	
	\begin{eqnarray}
	&&\mathrm{det}(\gamma_{mn})=\left\{\, X^\mu\ , X^\nu\,\right\}^2\\
	&&\left\{\, X^\mu\ , X^\nu\,\right\}_{\mathrm{PB}}\equiv 
	\epsilon^{mn}\,\partial_m\,
	X^\mu\, \partial_n\, X^\nu\, 
	\end{eqnarray}

	The action (\ref{fix}) is reparametrization invariant provided the 
	auxiliary field $\Phi(\, \sigma\,)$ transforms as a world-sheet scalar 
	density:
	 
	 \begin{equation}
	\Phi(\, \sigma^\prime\,)=\left\vert {\partial \sigma\over
	\partial\sigma^\prime\, }\,\right\vert\Phi(\, \sigma\,)
	\end{equation}
	
	Thus, by implementing reparametrization invariance
	 $\Phi(\,\sigma\,)$ can be transformed to unity and the 
	Schild action can be recovered as a ``gauge fixed'' form
	of (\ref{fix}):
	
	\begin{equation}
	I\left[\, \Phi=1\ , X\,\right]=\frac{\mu_0}{2}\, 
	\int_\Sigma d^2\sigma^\prime\,
	\left[\, det(\gamma_{mn}) 
	 + 1\,\right]=\mathrm{Schild} + \mathrm{const.}
	\label{schild}
	\end{equation}
	
	where, the numerical constant is proportional to
	the area of the integration domain $\Sigma$.\\
	On the other hand, by solving $\Phi(\, \sigma\,)$ in terms of $X$
	from (\ref{fix}) one recovers, ``on-shell'', the Nambu--Goto action
	
	\begin{equation}
	{\delta I\over \delta \Phi}=0\rightarrow \phi=\sqrt{\,
	det(\gamma_{mn})}\rightarrow I=\mu_0\,\int_\Sigma
	d^2\sigma\,\sqrt{-det(\gamma_{mn})}
	\label{ng}
	\end{equation}
	
	The inverse equivalence relation can be proven by starting from the
	Schild action \cite{schild}
	
	\begin{equation}
	I_S\equiv \mu_0\, \int_\Xi d^2\varphi\, \mathrm{det}\left[\, 
	\gamma_{ab}(\varphi)
	\,\right] 
	\end{equation}

	and  ``\textit{lifting}'' the original world--sheet coordinates
	$\varphi^m$ to the role of dynamical variables by mean of 
	reparametrization $\varphi^m\to \sigma^m=\sigma^m(\varphi)$ \cite{rep}:
	
	\begin{equation}
	I_{rep}\equiv \mu_0\, \int_\Sigma d^2\sigma\, \Phi^{-1}
	\mathrm{det}\left[\, \gamma_{ab}(\sigma)\,\right] \ ,\qquad
	\Phi^{-1}\equiv\epsilon_{ij}\,\epsilon^{mn}\,\partial_m\,
	\phi^i\, \partial_n\, \phi^j
	\label{irep}
	\end{equation}
	
	By variation of $I_{rep}$ with respect to $\varphi^i$ one gets
	the field  equation 
	
	\begin{equation}
	\epsilon_{ij}\,\epsilon^{mn}  \partial_n\, \phi^j \partial_m\,\left(\,
	{ det\left[\, \gamma_{ab}(\sigma)\,\right] \over \Phi^2}\,\right)=0
	\end{equation}
	
	Thus,
	\begin{equation}
	{det\left[\, \gamma_{ab}(\sigma)\,\right] \over \Phi^2}=
	\mathrm{const.}\equiv
	{1\over 4\mu_0}\label{undici}
	\end{equation}
	
	and the Nambu--Goto action (\ref{ng}) is ained again.\\
	This second option introduces a scalar doublet $\phi^i$, $i=1\ ,2$
	and  expresses the scalar density $\Phi$ as a ``\textit{composite}'' 
	object, rather than a fundamental one, or as a second
	``\textit{integration measure}'' \cite{guend}. 
	In any case the final result is unchanged.\\
	The action (\ref{fix}) is a special case of the general 
	two-parameter family of p-brane actions \cite{oda}
	
	\begin{equation}
	I^p_n\equiv \frac{\mu_0^{(p+1)/2}}{n}\, \int_\Sigma
	d^{p+1}\sigma\,e(\,\sigma\,)\,
	\left[\, {(\mathrm{det}\gamma_{mn})^{n/2}\over 
	 e(\,\sigma\,)^n} + n-1\,\right]
	\label{np}
	\end{equation}
	
	 where, in our notation $ e(\,\sigma\,)$ has been replaced
	 by $\Phi(\,\sigma\,)$ and $\mathrm{det}\gamma_{mn}$  is now
	 the square of the Nambu-Poisson bracket $\left\{\, X^{\mu_1}\ ,
	 \dots\ ,X^{\mu_{p+1}}\,\right\}_{NPB}$.\\ 
	 For $n=2$ and $p=1$ we obtain (\ref{fix}), while for $n=1$
	 the auxiliary field decouples and we get the Nambu--Goto action.\\
	In a naive, but not too much, way of thinking, one could trace back
	the correspondence between Yang--Mills gauge fields and Schild strings
	to the common quadratic form of both actions in the respective sets
	of ``field strengths'':
	
	\begin{equation}
	{1\over 4}\,\left\{\, X^\mu\ , X^\nu\,\right\}_{\mathrm{PB}}
	\left\{\, X_\mu\ , X_\nu\,\right\}_{\mathrm{PB}}
	\longleftrightarrow {1\over 4}\, \mathrm{Tr}\mathbf{F}_{\mu\nu}\,
	 \mathbf{F}^{\mu\nu}\label{analogy}            
	\end{equation}
	
	Pushing this formal analogy a little forward, and taking into account
	a possible gauge field type formulation of string dynamics \cite{gauge},
	one would expect
	to find a similar relation between the Nambu--Goto action and
	a non-Abelian Born--Infeld type action
	
	\begin{equation}
	\sqrt{\,\left\{\, X^\mu\ , X^\nu\,\right\}_{\mathrm{PB}}
	\left\{\, X_\mu\ , X_\nu\,\right\}_{\mathrm{PB}}\,}
	\longleftarrow \quad (?)\quad\longrightarrow
	\sqrt{\, \mathrm{Tr}\mathbf{F}_{\mu\nu}\,
	 \mathbf{F}^{\mu\nu}\,} \label{analogy2}         
	\end{equation}
	
	While being suggestive, relation (\ref{analogy2}) suffers from
	various problems not present in (\ref{analogy}), e.g.
	the very definition of the non-Abelian version of the
	Born-Infeld action is ambiguous \cite{tseit} .
	As the relation (\ref{analogy}) can be obtained through
	several, non-trivial steps, including ``quenching'', large--$N$
	expansion, Wigner-Weyl-Moyal quantization,
	it is the purpose of this communication to investigate how
	this approach can be, eventually, extended to the square root type 
	gauge action in (\ref{analogy2}). \\
	The non-perturbative aspects of the Yang-Mills models are better 
	described by transforming the original 
	gauge field theory into a {\it Matrix Quantum Mechanics.}  Such a
	transition is realized through dimensional reduction and quenching.
	The technical steps which allows to ``get rid of'' the internal,
	non-Abelian,
 indices $i$, $j$   and replace the spacetime coordinates $x^\mu$ with
 two continuous coordinates $\left(\, \sigma^0\ ,\sigma^1\,\right)$,
	are described in some detail elsewhere \cite{noi1},
	and will not be repeated here. For the reader convenience we shall
	give only sketch the main steps. The general procedure can be summarized
	as follows. \\
	Take the large--$N$ limit, i.e. let the row and column labels
	$i$, $j$ to range over arbitrarily large values. Thus,
	 $SU(  N )\to U(  N )$ and the group of spacetime
translations fits into the diagonal part of $U( \infty )$. By neglecting
off-diagonal components, spacetime dependent dynamical variables can be
shifted to the origin by means of a translation operator $\mathbf{U}(x)$:
since the translation group is Abelian one can choose the matrix $\mathbf{U}(x)$
to be a plane wave diagonal matrix \cite{kita}
\begin{equation}
\mathbf{U}_{ab}(x)=\delta_{ab}\exp\left(  i  q^a{}_\mu  x^\mu \right)
\ ,
\end{equation}
where $ q^a {}_\mu $ are the eigenvalues of the four-momentum $\mathbf{q }_\mu$.
Then
$$
\mathbf{A}_\mu(x)=\exp\left(-i\mathbf{q }_\mu  x^\mu \right)  \mathbf{A}_\mu(0)
\exp\left(  i\mathbf{q }_\mu  x^\mu \right)\equiv
\mathbf{U}^\dagger(x)  \mathbf{A}_\mu^{(0)} \mathbf{U}(x)
$$
and in view of the equality
$$
\mathbf{D}_ \mu  \mathbf{A}_\nu = i \mathbf{U}^\dagger(x)
 \left[  \mathbf{q }_\mu+\mathbf{A}_\mu^{(0)} ,
\mathbf{A}_\nu \right] \mathbf{U}(x),
$$
which when antisymmetrized yields
$$
\mathbf{D}_{[  \mu} \mathbf{A}_{\nu ]} = i \mathbf{U}^\dagger(x)
\left[  \mathbf{q }_\mu +\mathbf{A}_\mu^{(0)} ,
  \mathbf{q }_\nu  +  \mathbf{A}_\nu^{(0)} \right]\mathbf{U}(x)\nonumber\\
\equiv i   \mathbf{U}^\dagger(x)  \left[  \mathbf{A}_\mu^{(\mathrm{q})} ,
\mathbf{A}_\nu^{(\mathrm{q})} \right] \mathbf{U}(x)
,
$$
we can see that the translation is compatible with the
covariant differentiation, so that
$$
\mathbf{F}_{\mu\nu}(x)= \exp\left(-i\mathbf{q }_\mu x^\mu \right)  \mathbf{F
}_{\mu\nu}(0)\exp\left(  i\mathbf{q }_\mu x^\mu  \right)\equiv
\mathbf{U}^\dagger(x)  \mathbf{F}_{\mu\nu}^{(0)}  \mathbf{U}(x)
\ .
$$

\textit{Quenching Approximation} amounts to take into account the contributions
of the \textit{slow modes,} described by the eigenvalues of the momentum matrix
$\mathbf{q }$ and ``integrate out'' the non-diagonal fast modes. The final
result is to turn the original gauge theory is transformed into a quantum
mechanical model where the physical degrees of freedom are carried by
large coordinate independent matrices. Finally, 
the spacetime volume integration is regularized by enclosing the system
 in a ``box'' of  four-volume $V$
$$
\int d^4x\quad\longrightarrow \quad V
$$
In a Yang-Mills framework one can relate $V$ to the $QCD$ scale, i.e.
  $V=\left(\, 2\pi/\Lambda_{\mathrm{QCD}}\,\right)^4$ \cite{qcd}. Here,
  we shall determined $V$ by matching the large--$N$ limit of our matrix model
  with the Nambu--Goto action.\\

The resulting quenched action is

\begin{equation}
S^{(\mathrm{q})}_{\mathrm{YM/\Phi }}=  
   {NV\over 4g_0^2 }  \mathrm{Tr}
\left(  \left[  \mathbf{A}_\mu^{(\mathrm{q})} , 
\mathbf{A}_\nu^{(\mathrm{q})} \right]^2
\right)\ .
\label{sikkt}
\end{equation}

The large--$N$  quantum properties of the matrix model (\ref{sikkt}) can be 
effectively investigated by means of the Wigner-Weyl-Moyal correspondence
between matrices and functions, i.e. \textit{Symbols}, defined over a
\textit{noncommutative} phase space.  
The resulting theory is a {\it
deformation} of an ordinary field theory, where the ordinary product
between functions is replaced by a non-commutative $\ast$--product.
The deformation parameter, measuring the amount of non-commutativity,
results to be $1/N$, and the classical limit corresponds exactly to the
large--$N$ limit. The final result is a string action of the Schild type, 
which is invariant under area-preserving reparametrization of the world-sheet.\\
More recently, we have also shown that \textit{bag\/}-like objects fit the 
large--$N$ spectrum of Yang--Mills type theories as well, both in four 
\cite{noi1} and higher dimensions.\\
 Here, we would like to explore a different route leading in
a  straightforward way to a Nambu--Goto string action. Having discussed the
role of the auxiliary field $\Phi$ in bridging the gap between Schild and
Nambu--Goto actions at the classical level, we propose the following improved
gauge matrix model
	
	\begin{equation}
S^{(\mathrm{q})}_{\mathrm{YM/\Phi\,  }}=  
   {N\, V\over 4g_0^2 }  \mathrm{Tr}\left[\,  \mathbf{\Phi}^{-1}
\left(  \left[  \mathbf{A}_\mu^{(\mathrm{q})} , 
\mathbf{A}_\nu^{(\mathrm{q})} \right]^2
\right)\,\right] +{1\over 4}\,\mathrm{Tr}\mathbf{\Phi}
\ .
\label{improved}
\end{equation}
	
where, $\mathbf{\Phi}$ is a $ N\times N $ diagonal matrix. In a different
framework, a similar matrix variable have been introduced to build a consistent
path integral for the matrix version of type IIB superstring model
\cite{makeenko}.\\
We see, as things stand up to now there seems to be a qualitative
difference between equation 
(\ref{improved}) and equation (\ref{irep}). This is the absence of the 
analogous to the last term in eq. (\ref{improved}), the one that contains only 
$\mathrm{Tr}(\Phi)$, which is absent in eq.(\ref{irep}).
 Nevertheless the corresponding demand that the "measure" $\Phi$ be a
``total derivative'' can be implemented in  simple ways also in the matrix
model. The first way to implement this is as follows: let us take
(\ref{improved}) with the first term in the 
right hand side only, but let us also say that the variation with respect
to $\Phi$ has to be performed taking into account that $\Phi$ is in some sense 
the anologous of a ``total derivative''. 
But then, what is a total derivative in the matrix formulation?
One property of a total derivative  is that its integral is
is fixed from the boundaries, which are not varied. Let us say for 
simplicity that we take the integral to be zero ( fixing it to another
constant will not change anything ).
In this case we must proceed as follows: 
\begin{enumerate}
\item Take the action as in (\ref{improved}),
but only consider the  first term in the right hand side.
\item Consider the variation of such action, but with the constraint that 
$\Phi$ is 
a total derivative, which means that  $\mathrm{Tr}(\Phi) = 0$ .
\item To do this in practice we add to the action defined in (\ref{fix})
$c\, Tr(\Phi)$, where $c$ is an undetermined lagrange multiplier
which implements condition $(2)$.
The resulting theory will have then an arbitrary string tension if we continue 
from $(2)$. The constant $c$ , i.e. the
undetermined lagrange multiplier playing the role of the 
constant of integration in (\ref{undici}).
\end{enumerate}
Yet another way of proceeding is to explicitly construct a composite
measure in the matrix formulation ( instead of using a property a total
derivative must satisfy and then implement this ).
The procedure in this case would be:
 \begin{itemize}
 \item
 i) the world sheet coordinates are replaced by matrices  $\xi^m$;
  \item
ii) the derivatives are replaced by commutators between
the matrix coordinates and the matrix momentum  $q_a$.
\end{itemize}
The end result is a matrix $\Phi$ :
$\Phi = \epsilon^{ab} \epsilon_{mn}\,\left[\, q_a\ ,\xi^m \,\right]
\left[\, q_b\ ,\xi^n\,\right]$
which should be used in (\ref{improved}), except that now the second term 
in (\ref{improved}) and/or a possible lagrange multiplier  is now
not necesary ( the constraint $\mathrm{Tr}(\Phi) = 0$ is now an identity).
Now, going back to the action (\ref{improved}),
if $\mathbf{\Phi}$ is taken to be the $N\times N$ \textsl{Identity} matrix,
then action (\ref{improved}) takes on the form of the Yang-Mills action. On the
other hand, if we let $\mathbf{\Phi}$ to be diagonal, 
$\Phi_{ab}\equiv \phi_{(a)}\, \delta_{ab}$ ( no summation over the index $a$ ),
and determine its eigenvalues by varying (\ref{improved}), we find

\begin{equation}
\phi_{(a)}=\sqrt{{N\, V\over g_0^2}\, \mathbf{F^{(0)}}{}_{cd\, \mu\nu} 
\mathbf{F}^{(0)}{}^{cd\,\mu\nu} }\equiv \sqrt{\, {N\,V\over g_0^2}\,
\mathrm{Tr}\,\mathbf{F^{(0)}}{}_{\mu\nu} \mathbf{F}^{(0)}{}^{\mu\nu} }
\label{eigen}
\end{equation}

All the eigenvalues are degenerate and quadratic in the Yang-Mills field
strength. By inserting  (\ref{eigen}) into (\ref{improved}) one finds:

\begin{equation}
S^{(\mathrm{q})}_{\mathrm{BI-YM}}= 
 \sqrt{\, {N\, V\over 4g_0^2 }\,  
\mathbf{F}_{cd\, \mu\nu}^{(0)} \,\mathbf{F}^{(0)}{}^{cd\,\mu\nu}\,}\equiv
 \sqrt{\, {N\, V\over 4g_0^2 } \mathrm{Tr}\, 
\mathbf{F}_{\mu\nu}^{(0)}\, \mathbf{F}^{(0)}{}^{\mu\nu}\,}
\label{nonabi}
\end{equation}

The action (\ref{nonabi}) is the ``square root'' form of a non-Abelian
Born-Infeld type action, where the trace is the standard one. The
ambiguity in the definition of the trace over internal indices is removed,
in our model, by choosing $\mathbf{\Phi}$ to be diagonal. With hindsight,
we know that in the large--$N$ limit the trace over internal $SU(\,N\,)$
indices will turn into an integration over world manifold coordinates.
Thus, it is compelling to ``move out'' the trace from the square root, in
order to obtain a Nambu--Goto type action integral. What we are going to 
describe is a procedure such that in the  large--$N$ limit:

\begin{equation}
\sqrt{\, \mathbf{Tr}\left(\, \dots\, \right)}\longrightarrow 
\int_\Sigma d^2\sigma \sqrt{\, \dots\, }
\end{equation}

and the non-commuting  Yang--Mills matrices are replaced by commuting string 
coordinates. \\
 The string world--sheet is the target spacetime  image
$x^\mu=X^\mu(\,  \sigma^0,\sigma^1\,)$ of the world
manifold $\Sigma$ :  $-\infty\le \sigma^0\le +\infty$, $0\le \sigma^1\le L$,
$X^\mu$ belonging to the algebra $\mathcal{A}$ of $C^\infty$, of functions
 over $\Sigma$. Thus,
  to realize our program we must \textit{deform} ${\mathcal{A}}$ to a
\textit{non-commutative ``starred'' algebra} by introducing a $\ast$-product.
The general rule is to define  the new product between two
functions  as  (for a recent review see \cite{harvey}):
\begin{equation}
f \ast  g =f \,g + \hbar\,  P_\hbar(\,  f\ ,g \,)
\ ,
\end{equation}
where $P_\hbar(\,  f\ ,g\, )$ is a bilinear map
$P_\hbar : {\mathcal{A}}\times {\mathcal{A}}\rightarrow
{\mathcal{A}}$. $\hbar$ is the \textit{deformation parameter,} which,
in our case, is defined as $ \hbar\equiv {2\pi/ N} $.
The Moyal product is defined as the deformed $\ast$-product
\begin{equation}
    f(\sigma)  \ast   g(\sigma)
    \equiv
    \exp\left[  i  {\hbar\over 2} \omega^{mn}
    {\partial^2\over \partial  \sigma^m   \partial \xi^n}
    \right] f(\sigma)  g(\xi)
    \Biggr\rceil_{\xi=\sigma}
  \ ,
    \label{mp}
\end{equation}
	
	where $\omega^{mn}$ is a non-degenerate, antisymmetric matrix, which
can be locally written as
\begin{equation}
\omega^{mn}=\left( \begin{array}{cc}
                 0  & 1\\
                -1 & 0
                \end{array}
                  \right)
               \ .
\end{equation}
The Moyal product (\ref{mp}) takes a simple looking form in Fourier space
\begin{equation}
F(\sigma)  \ast   G(\sigma)=\int {d^{  2}\xi\over (2\pi)}
\exp\left(  i  {\hbar\over 2} \omega_{mn}
\sigma^m    \xi^n \right)
F\left(  {\sigma \over 2}+ \xi  \right)
G\left(  {\sigma  \over 2}- \xi  \right)
\ ,
\end{equation}
where $F$ and $G$ are the Fourier transform of $f$ and $g$.
Let us consider the Heisenberg algebra
\begin{equation}
\left[\,  \mathbf{K}\ , \mathbf{P}\, \right]=i\, \hbar \ ;
\end{equation}
Weyl suggested, many years ago, how an operator  ${\mathbf{O}}_F( \mathbf{K},
\mathbf{P} )$ can be written as a sum of  algebra elements as
\begin{equation}
\mathbf{O}_F= {1\over (2\pi)}\int dp \, dk\,  F\left(\,  p\ ,k \,\right)
\exp\left(  i  p \,\mathbf{K} + i   k \mathbf{P} \right)
\ .
\label{weyltrasf}
\end{equation}
The Weyl map (\ref{weyltrasf}) can be inverted to associate functions, or
more exactly \textit{symbols}, to operators
\begin{equation}
F\left(\,  q\ ,k \,\right)= \int {d\xi\over (2\pi)} \exp\left(-i k \xi \right)
\Biggl\langle   q + \hbar{\xi \over 2}\Biggl\vert \mathbf{O}_F\left(\,\mathbf{K}\ 
,
 \mathbf{P} \,\right)  \Biggr\vert   q - \hbar {\xi \over 2 }  \Biggr\rangle
\ ;
\label{invweyl}
\end{equation}
moreover it translates the commutator between two
operators ${\mbox{\boldmath{$U$}}}$, ${\mbox{\boldmath{$V$}}}$ into the
\textit{Moyal Bracket} between their
corresponding symbols  ${\mathcal{U}}(\sigma)$, ${\mathcal{V}}(\sigma)$
\[
    {1\over i \hbar}   \left[   {\mbox{\boldmath{$U$}}},
    {\mbox{\boldmath{$V$}}}  \right]\longleftrightarrow
    \left\{{\mathcal{U}}, {\mathcal{V}}\right\}_{\mathrm{MB}}
    \equiv
    {1\over i  \hbar}
    \left(
    {\mathcal{U}} \ast {\mathcal{V}} - {\mathcal{V}} \ast {\mathcal{U}}
      \right)
\]
and the quantum mechanical trace into an integral over Fourier space.
A concise but pedagogical introduction to
the deformed differential calculus and its application to
the theory of integrable system can be found in \cite{strach}.
	We are now ready to formulate the alleged relationship between the
	quenched
model (\ref{nonabi}) and string model: the symbol of the matrix
$\mathbf{A}^{(\mathrm{q})}_\mu$ is proportional to the string coordinates
$X^\mu( \, \sigma^0\ , \sigma^{1}\, )$. Going through the steps discussed
above the action $S^{(\mathrm{q})}_{\mathrm{YM/\Phi }}  $ transforms into its symbol
$W^{(\mathrm{q})}_{\mathrm{YM/\Phi }}  $:
\begin{equation}
S^{(\mathrm{q})}_{\mathrm{YM/\Phi }}\to
W^{(\mathrm{q})}_{\mathrm{YM/\Phi }}=
{N  V\over 8\pi   g_0{}^2 }
\int_\Sigma d^{2}\sigma\left[\, {1\over \Phi(\sigma)}\ast
 \left\{\,  A_\mu(\sigma)\ , 
\mathbf{A}_\nu(\sigma) \,\right\}_{\mathrm{MB}}^2\, +{g_0{}^2\over N\,V\, }\,
\Phi(\sigma)\,\right]\ .
\label{wbfss}
\end{equation}

and we rescale the Yang--Mills charge and field\footnote{For the sake of
clarity, let us summarize the canonical dimensions in natural units of
various quantities:
\begin{eqnarray}
  &&\left[\, A_\mu{}^a(x)\, \right] \equiv
[\, \mathbf{A}_\mu^{(\mathrm{q})}\, ]=(  \mathrm{length} )^{-1}\ ,\qquad
[\,  F_{\mu\nu}{}^a(x)\, ] \equiv
[\, \mathbf{F}_{\mu\nu}{}^{(\mathrm{q})}\,]=(  \mathrm{length} )^{-2}\nonumber\\
&& 
[ \, g_0\, ]\equiv [ \, g\, ]=(  \mathrm{length} )^0=1\ ,\qquad
[\,  V\, ]= (  \mathrm{length} )^4  \ ,\qquad
 [\,  \mu_0\, ]= (  \mathrm{length} )^{-2}
 \ .
\nonumber
\end{eqnarray} 
}
as
\begin{eqnarray}
{N \over g_0{}^2} && \quad \longmapsto \quad {1 \over g^2}\ ,\qquad \mathrm{Tr}
\longmapsto \frac{1}{2\pi\sqrt V}\int_\Sigma d^2\sigma
\\
 A^\mu && \quad \longmapsto \quad V^{-1/2}\,  X^\mu\ ,\qquad
 \mathbf{F}^{(0)}{}^{\mu\nu}\longmapsto \quad V^{-1/2}\left\{  X^\mu\ , X^\nu
 \right\}_{\mathrm{MB}}
\ .
\end{eqnarray}

	Finally, if $N \gg 1$ the Moyal bracket can be approximated by the
	Poisson bracket
$$
\left\{  X^\mu\ , X^\nu \right\}_{\mathrm{MB}}
\longmapsto
\left\{  X^\mu\ , X^\nu \right\}_{\mathrm{PB}}
$$
and (\ref{wbfss}) takes the form 
\begin{equation}
W^{(\mathrm{q})}_{\mathrm{YM/\Phi }}\to S_{\mathrm{NG}}=
{1\over 8\pi   g^2\, \sqrt V }
\int_\Sigma d^{2}\sigma\left[\, {1\over \Phi(\sigma)}
\left\{\,  X_\mu(\sigma)\ , X_\nu(\sigma) \,\right\}_{\mathrm{PB}}^2 
+ g^2 \,\Phi(\sigma)\,\right]
\label{nbi}
\end{equation}

which is (\ref{fix}) provided we identify 

\begin{equation}
\mu_0\longrightarrow\equiv {1\over 4\pi\,   g\, \sqrt V }	
\end{equation}	
	
According with the initial discussion we can establish the following, large--$N$,
correspondence:

\begin{equation}
S^{(\mathrm{q})}_{\mathrm{BI-YM}}\approx
 S^{(\mathrm{q})}_{\mathrm{YM/\Phi  }} \to {1\over 4\pi\,   g\, \sqrt V
 }\int_\Sigma d^2\sigma \,\sqrt{-det(\gamma_{mn})}
\end{equation}

As a concluding remark, it may be worth mentioning that the approach discussed
above can be extended in a straightforward way to a more general non-Abelian
Born-Infeld action including topological terms. Instead of starting
from (\ref{improved}), one can consider

	\begin{eqnarray}
S^{(\mathrm{q})}_{\mathrm{YM/\Phi\,  }}=&& \frac{M^4\, V}{2}
     \mathrm{Tr}\left\{\,  \mathbf{\Phi}^{-1}
\left[\, \mathbf{I} +{N\over 4g_0^2 M^4 }\mathbf{F}_{\mu\nu}\,
 \mathbf{F}_{\mu\nu}\right.\right. \nonumber\\ 
 +&& \left. \left. \left(\, {N\over 4g_0^2 M^4 }\,\right)^2
 \,\left(\,\epsilon^{\lambda\mu\nu\rho}\, \mathbf{F}_{\lambda\mu}\,
 \mathbf{F}_{\nu\rho}\,\right)^2
\right]- \mathbf{I} \right\}
  +{M^4\,V\over 2}\,\mathrm{Tr}\mathbf{\Phi}
\ .
\label{improved2}
\end{eqnarray}
where, $\mathbf{I}$ is the $N\times N $ identity matrix and $M$ a new mass
scale. The coefficient in front of the topological density has been assigned
in analogy to the Abelian case, where no trace ambiguity occurs and the square
root form is derived by expanding the determinant of the matrix
$M_{\mu\nu}\equiv\delta_{\mu\nu} +\mathrm{const.}\times F_{\mu\nu}$.\\
The topological density contribution trivially vanishes in the large--$N$
limit:

\begin{equation}
\epsilon^{\lambda\mu\nu\rho}\, \mathbf{F}_{\lambda\mu}\,
 \mathbf{F}_{\nu\rho}\longrightarrow 
\epsilon^{\lambda\mu\nu\rho}\,\left\{\,  X_\lambda(\sigma)\ , X_\mu(\sigma)
\,\right\}_{\mathrm{PB}}\left\{\,  X_\nu(\sigma)\ , X_\rho(\sigma)
\,\right\}_{\mathrm{PB}}\equiv 0
\end{equation}
	
A non-vanishing contribution from the topological term can only be obtained
if strings are replaced by higher dimensional extended objects \cite{noi1}.
Thus, as far as strings are concerned, one finds

\begin{equation}
S^{(\mathrm{q})}_{\mathrm{YM/\Phi\,  }}\to S^{(\mathrm{q})}_{\mathrm{BI\, 
}}=\frac{M^4\,\sqrt V}{2\pi}\,
\int_\Sigma d^2\sigma \,\left[\, \sqrt{ \left(\,  
1 + \frac{1}{4g^2\, M^4\,  V }\,\mathrm{det}[\,\gamma_{mn}\,]\,\right)} 
-1\,\right]\label{bis}
\end{equation}

Recent results about $UV/IR$ interplay, in the framework of noncommutative 
Yang-Mills theories \cite{szabo}, suggest to investigate the behavior of
the action (\ref{bis}) by changing the size of the quantization volume $V$.
In the  \textit{small volume} limit, i.e. $ V  << 1/ 4g^2\, M^4 $, the main
contribution to $S^{(\mathrm{q})}_{\mathrm{BI\, }}$ comes from the
$\mathrm{det}[\,\gamma_{mn}\,]$ 
and the Nambu--Goto action is recovered again. The quantization volume
drops out and the string tension, i.e.  $\mu_0=M^2/4\pi g$, is determined
by the only relevant mass scale $M$. \\
In the opposite, \textit{large volume,} limit, i.e. $ V  >> 1/ 4g^2\, M^4 $,
the second term in the square root is small with respect to $1$ and the
first non-vanishing contribution of the Taylor expansion is the Schild
action. In this regime the relevant energy scale is set $V^{-1/4}$ and the
corresponding string tension is $\mu_0=1/32\pi g^2\sqrt V$.
\\
From a different point of view, Fairlie has recently pointed out some
intriguing analogy between the Born-Infeld  and  Nambu--Goto actions \cite{df}.
Our results supports this connection. We believe we 
 bridged the gap between four dimensional 
gauge theories and fully reparametrization invariant string models.
 We found new, non-trivial,
relationship between a class of generalized $SU(\, N\,)$ models and 
Born-Infeld/Nambu--Goto strings, provided a new matrix degree of freedom,
$\mathbf{\Phi}$, is  introduced. In the original $SU(\, N\,)$ model  
$\mathbf{\Phi}$ connects the \textit{Yang-Mills phase,} at $\mathbf{\Phi}=
\mathbf{I}$, with the, non-linear, \textit{Born-Infeld phase,} where the
eigenvalues $\phi_{(a)}$ are given by equation (\ref{eigen}). 
In the large--$N$ limit
the order parameter becomes a world-sheet auxiliary field linearizing a
square root type Born-Infeld string model. The small, respectively,  large  
volume limits of this model correspond to  Nambu--Goto and Schild actions.
	


\begin{thebibliography}{99}
	\bibitem{uno} I. Bars  
	Phys. Lett.  \textbf{245B} 35   (1990)\\
    E.G. Floratos, J. Illiopulos and G. Tiktopoulos 
    Phys. Lett.  \textbf{B 217}  285   (1989) \\
    D. Fairlie and C.K. Zachos 
   Phys. Lett.  \textbf{B224} 101 (1989)
	\bibitem{oda} I. Oda
	Chaos, Sol. \& Fract. \textbf{10}, n.2/3,   483 (1999)\\
	 S. Ansoldi,  C. Castro, E. Spallucci
	 Class. Quant. Grav. \textbf{L97} (2001)
	\bibitem{schild} A. Schild 
	Phys. Rev. D\textbf{16} 1722 (1977)
	\bibitem{rep} S. Ansoldi, A. Aurilia, E. Spallucci
	Phys. Rev. D\textbf{53} 870 (1996)
	\bibitem{guend} E. I. Guendelman
       Class. Quant. Grav. \textbf{17}  3673 (2000)\\
	 E. I. Guendelman
     Phys. Rev. D \textbf{63}  046006  (2001 )
	\bibitem{gauge} A. Aurilia, A. Smailagic, E. Spallucci
	 Phys. Rev.  D\textbf{47}  2536 (1993)
	\bibitem{tseit} T. Hagiwara, J. Phys. A\textbf{14},  3059 (1981)\\
	A. A. Tseytlin 
	Nucl. Phys. B\textbf{501}  41   (1997) \\
	J. H. Park Phys. Lett. B \textbf{458}  471 (1999)
	\bibitem{kita} Y. Kitazawa,S.R.  Wadia 
         Phys. Lett.  \textbf{120B} 377 (1983)
	\bibitem{noi1}  S. Ansoldi ,  C. Castro, E. Spallucci
    Phys. Lett. \textbf{B504} 174   (2001) \\
	 S. Ansoldi ,  C. Castro, E. Spallucci
	  Class. Quant. Grav. \textbf{18}  L17 (2001)
	\bibitem{makeenko} A. Fayyazuddin, Y. Makeenko, P. Olesen, D.J. Smith,
	K. Zarembo
	Nucl. Phys. \textbf{B499} 159 (1997)\\
	for a review see Y. Makeenko 
    \textit{Three Introductory Lectures in Helsinki on Matrix Models of
    Superstrings, } hep-th/9704075 (1997)
	\bibitem{qcd} S. Ansoldi, C. Castro, E.Spallucci
	Class. Quantum Grav. \textbf{18}  2865 (2001)
	\bibitem{harvey} Harvey J A 
\textit{Komaba Lectures on Noncommutative Solitons and D-Branes,} hep-th/0102076
(2001)
\bibitem{strach} I. A. B. Strachan 
    \textit{J. Geom. and  Phys.} \textbf{21}  255 (1997)
	\bibitem{df} D. B. Fairlie
	Phys. Lett. \textbf{B456},  141 (1999)
	\bibitem{szabo}  R. J. Szabo
	\textit{Quantum Field Theory on Noncommutative Spaces,}
	hep-th/0109162 (2001)
	\end{thebibliography}
	\end{document}